\documentstyle[12pt]{article}

\newlength{\dinwidth}
\newlength{\dinmargin}
\newcommand{\resection}[1]{\setcounter{equation}{0}\section{#1}}

\begin{document}
\vspace*{4cm}
\begin{center}
  \begin{Large}
  \begin{bf}
AN EXTENSION OF THE ELECTROWEAK MODEL  WITH DECOUPLING\\ AT LOW ENERGY\\
  \end{bf}
  \end{Large}
  \vspace{5mm}
  \begin{large}
R. Casalbuoni,  S. De Curtis and D. Dominici\\
  \end{large}
Dipartimento di Fisica, Univ. di Firenze\\
I.N.F.N., Sezione di Firenze\\
  \vspace{5mm}
  \begin{large}
M. Grazzini\\
  \end{large}
Dipartimento di Fisica, Univ. di Parma \\
I.N.F.N., Gruppo collegato di Parma
  \vspace{5mm}
  \vspace{5mm}
\end{center}
  \vspace{2cm}
\begin{center}
University of Florence - DFF 251/6/96 
\end{center}
\vspace{1cm}
\noindent
\newpage
\thispagestyle{empty}
\begin{quotation}
\vspace*{5cm}

\begin{center}
\begin{bf} 
  ABSTRACT
  \end{bf}
\end{center}
\vspace{1cm}
\noindent
We present a renormalizable model of electroweak interactions containing an 
extra 
$SU(2)'_L\otimes SU(2)'_R$ symmetry. The masses of the corresponding 
gauge bosons and of
the associated Higgs particles can be made heavy by tuning a convenient
vacuum expectation value. According to the way in which the heavy mass 
limit is
taken we obtain a previously considered non-linear model (degenerate BESS)
which, in this
limit, decouples giving rise to the Higgsless Standard Model (SM). 
Otherwise we can get a
model which decouples giving the full SM. In this paper we argue that in
the second limit the decoupling holds true also at the level of
radiative corrections. Therefore the model discussed here is not
distinguishable from the SM at low energy. Of course the two models differ
deeply at higher energies.

  \vspace{5mm}
\noindent
\end{quotation}
\newpage
\setcounter{page}{1}
\def\spin{{2}}
\def\uu{\dd{u^2}}
\def\vv{\dd{v^2}}   
\def\lq{\left [}
\def\rq{\right ]}
\def\qq{Q^2}
\def\ct{c_{\theta}}
\def\st{s_{\theta}}
\def\cpsi{\cos\psi}
\def\spsi{\sin\psi}
\def\sf{s_{\varphi}}
\def\cf{c_{\varphi}}
\def\dmus{\partial^{\mu}}
\def\dmu{\partial_{\mu}}
\def\LL{{\cal L}}
\def\BB{{\cal B}}
\def\Tr{{\rm Tr}}
\def\gp{g'}
\def\gs{g''}
\def\ggs{\frac{g}{\gs}}
\def\eps{{\epsilon}}
\def\f{\frac}
\def\L{{W^{\prime }}_L}
\def\R{{W^{\prime }}_R}
\newcommand{\be}{\begin{equation}}
\newcommand{\ee}{\end{equation}}
\newcommand{\bea}{\begin{eqnarray}}
\newcommand{\eea}{\end{eqnarray}}
\newcommand{\nn}{\nonumber}
\newcommand{\dd}{\displaystyle}

\resection{Introduction}

In a previous paper \cite{degene} we have considered a model 
(degenerate BESS)
of electroweak
interactions describing, besides the usual $W^\pm$, $Z$ and $\gamma$ vector
bosons, two new triplets of spin 1 particles, $V_ L$ and $V_R$. 
These new states are degenerate in mass if one neglects their mixing
to the ordinary vector bosons. The description of the model was based on a
non-linear gauged $\sigma$-model and we refer to \cite{degene} for more
details.
The interest in the model was due to its decoupling properties: in the limit
of infinite
mass of the heavy vector bosons  one gets back the SM.
This is a rather non trivial property because one is dealing with a non-linear
theory with couplings increasing with the heavy masses. In fact, the 
decoupling
originates from an accidental global symmetry that the model possesses when 
the
gauge couplings are turned off. This is also the symmetry from which the 
quasi-degeneracy of the heavy vector states arises.

It is an interesting question by itself to ask if a linear version of the 
model
does exist. The original philosophy underlying the non-linear version was 
based
on the idea that the non-linear realization would be the low-energy 
description
of some underlying dynamics giving rise to the breaking of the electroweak
symmetry. In this respect looking for a linear realization might appear as 
based
on a completely different standpoint. However we are thinking of a scenario 
very close
to the one arising in non-commuting technicolor models \cite{nctc},
where one has an underlying strong dynamics producing heavy Higgs composite
particles. In this sense we are trying to describe the theory at the level of
its composite states, vectors (the new heavy bosons), and scalars (Higgs 
bosons).
That is, we are looking at a scale in which the Higgs bosons are yet relevant 
degrees 
of freedom. The advantage of this is that of  dealing with a renormalizable 
theory. 
By that one is able to discuss if the decoupling holds at the level of
radiative corrections. We will argue that the linear realization of the model
decouples, and
consequently that the high-energy physics we are talking about is not relevant
at the LEP scale.

In the following we will describe the linearized version of the model showing
that, at tree level, it coincides with the non-linear model of ref. 
\cite{degene}. 
We will then prove that, by diagonalizing the vector boson mass
matrices and via a redefinition of the gauge couplings, all the couplings in 
the light
and in the light-heavy sectors do
not increase with the heavy masses. From this
we can argue that the theory decouples also at the level of the radiative
corrections. A detailed check of this point by means of an explicit
calculation will be given in a more technical
and complete paper \cite{lavoro}. We can therefore say that the model we
present is identical to the standard model in its low energy manifestations,
although at higher energies the differences can be rather dramatic 
\cite{degene}.

\resection{The Model}
The model we consider here is based on a gauge group 
$SU(2)_L\otimes U(1)\otimes SU(2)_{L}^\prime \otimes SU(2)_{R}^\prime$
and has a scalar sector consisting of scalar fields belonging to the following 
representations 
of the group 
$SU(2)_L\otimes SU(2)_R\otimes SU(2)_{L}^\prime \otimes SU(2)_{R}^\prime$   
\be
\tilde L\in (\spin,0,\spin,0)~~~~~
\tilde U\in (\spin,\spin,0,0)~~~~~
\tilde R\in (0,\spin,0,\spin)
\ee
that is with transformation properties
\be
{\tilde L}^\prime = g_L\tilde  L h_L~~~~~
{\tilde U}^\prime = g_L\tilde U g_R^\dagger ~~~~
{\tilde R}^\prime = g_R\tilde R h_R
\ee
where
\bea
& & g_L\in {SU(2)_L}~~~~~
g_R\in {SU(2)_R}\nn\\
 & &h_L\in SU(2)_L^\prime~~~~~
h_R\in SU(2)_R^\prime
\eea
We will see that with this system of scalar fields it is possible to break the gauge
symmetries through the following chain 
\be
\matrix{
SU(2)_L\otimes U(1)\otimes SU(2)_{L}^\prime \otimes SU(2)_{R}^\prime\cr
\downarrow u\cr
SU(2)_{\rm weak}\otimes U(1)_Y\cr
\downarrow v\cr
U(1)_{\rm em}}
\ee
The two breakings are induced by the expectation values 
$\langle\tilde L\rangle=
\langle\tilde R\rangle=u$ and $\langle\tilde U\rangle=v$ respectively. 
The first two expectation 
values make the breaking $SU(2)_L\otimes SU(2)'_L\to SU(2)_{\rm weak}$ and  
$U(1)\otimes 
SU(2)'_R\to U(1)_Y$, whereas the second breaks in the standard way
$SU(2)_{\rm weak}\otimes U(1)_Y\to U(1)_{\rm em}$. In the following we will 
assume 
that the first breaking corresponds to a scale $u\gg v$. This chain of 
breaking 
is reminiscent of the one conjectured in non-commuting extended technicolor 
theories
(NCETC) \cite{nctc}, where one has
\be
\matrix{
G_{ETC}\otimes SU(2)_{\rm light}\otimes U(1)\cr
\downarrow f\cr
G_{TC}\otimes SU(2)_{\rm heavy}\otimes SU(2)_{\rm light}\otimes U(1)_Y\cr
\downarrow u\cr
G_{TC}\otimes SU(2)_{\rm weak}\otimes U(1)_Y\cr}
\ee
where $G_{ETC}$ is the extended technicolor gauge group, and $G_{TC}$ the 
technicolor
one. If we make the following identifications $SU(2)_L=SU(2)_{\rm light}$,
$SU(2)'_L=SU(2)_{\rm heavy}$, one can think of an extension of the NCETC 
schemes
such as
\be
\matrix{
G_{ETC}\otimes SU(2)_L\otimes U(1)\cr
\downarrow f\cr
G_{TC}\otimes SU(2)'_L\otimes SU(2)_L\otimes SU(2)'_R\otimes U(1)\cr}
\ee
After  the $G_{ETC}$ has been broken the chain proceeds as in eq. (2.4). The 
original NCETC scheme has been here modified in order to allow for the gauge 
particles
transforming under $SU(2)'_R$,  ensuring, together with the 
vector bosons from $SU(2)'_L$, the decoupling at low energy.

Proceeding in a completely standard way, we can build up 
covariant derivatives with respect to the 
local
$SU(2)_L\otimes U(1)\otimes SU(2)_{L}^\prime \otimes SU(2)_{R}^\prime$
\bea
& &D \tilde L =\partial \tilde L +
i g_0 \frac{{\vec \tau}}{2}\cdot{\vec W} \tilde L
-i g_2 \tilde L\frac{{\vec \tau}}{2}\cdot{{\vec V}_L}  \nn\\
& &D \tilde R =\partial\tilde R +i  g_1 \frac{{ \tau_3}}{2}{Y}\tilde R
-i g_3\tilde R\frac{{\vec \tau}}{2}\cdot{{\vec V}_R}  \nn\\
& &D\tilde U =\partial\tilde U +
i g_0 \frac{{\vec \tau}}{2}\cdot{\vec W} \tilde U   
- i  g_1\tilde U\frac{{ \tau_3}}{2}{Y} 
\eea
where ${\vec V}_L~({\vec V}_R)$ 
are the  gauge fields in $SU(2)_L^\prime ~
(SU(2)_R^\prime)$,   
 with the corresponding gauge couplings  $g_2$, and
$g_3$, whereas $g_0$, $g_1$, are the gauge couplings of the $SU(2)_L$ and 
$U(1)$
gauge groups respectively.

This model contains, besides the standard Higgs sector given
by the field $\tilde U$, the additional scalar fields $\tilde L$ and $\tilde 
R$.

The lagrangian for the kinetic terms of these scalar fields
is given by
\be
{\cal L}^h = \frac 1 4 \lq Tr (D_\mu \tilde U)^\dagger  (D^\mu\tilde U)
+Tr (D_\mu \tilde L)^\dagger  (D^\mu \tilde L)
+Tr (D_\mu \tilde R)^\dagger  (D^\mu \tilde R) \rq            
\ee
We have then to discuss the scalar potential which is supposed to
break  the original symmetry down to the $U(1)_{\rm em}$ group.
The most general potential invariant with respect to the group
 $SU(2)_L\otimes SU(2)_R\otimes SU(2)_{L}^\prime \otimes SU(2)_{R}^\prime$ 
 is given by
\bea
&V(\tilde U,\tilde L,\tilde R)=&\mu_1^2 Tr ({\tilde L}^\dagger \tilde L) 
+\frac { \lambda_1} {4}[ Tr ({\tilde L}^\dagger \tilde L)]^2
+\mu_2^2 Tr ({\tilde R}^\dagger \tilde R) 
+\frac {\lambda_2} {4} [Tr ({\tilde R}^\dagger \tilde R)]^2\nn\\
& &+m^2  Tr ({\tilde U}^\dagger \tilde U) 
+\frac h 4 [Tr ({\tilde U}^\dagger \tilde U)]^2        
+\frac {f_3} {2} Tr ({\tilde L^\dagger}\tilde L) 
Tr ({\tilde R^\dagger}\tilde R)\nn\\
&&+\frac {f_1}{2} Tr ({\tilde L^\dagger}\tilde L ) 
Tr ({\tilde U^\dagger}\tilde  U)
+\frac {f_2}{2} Tr ({\tilde R^\dagger}\tilde R) 
Tr ({\tilde U^\dagger}\tilde U)       
\eea

In the following we will also require, for the scalar potential,
the discrete symmetry $L\leftrightarrow R$, which 
implies
\bea
&&g_3=g_2\nn\\
&&\mu_1=\mu_2=\mu\nn\\
&&\lambda_1=\lambda_2=\lambda\nn\\
&&f_1=f_2=f
\eea

The total lagrangian is obtained by adding the kinetic terms for
the gauge fields:
\be
{\cal L}={\cal L}^h-V(\tilde U,\tilde L,\tilde R)
+{\cal L}^{kin}(W,Y,V_L,V_R)
\ee
where
\bea
{\cal L}^{kin}(W,Y,V_L,V_R)&=
&\frac 1 2 {\rm tr}[F_{\mu\nu}(W)F^{\mu\nu}(W)]+
\frac 1 2 {\rm tr}[F_{\mu\nu}(Y)F^{\mu\nu}(Y)]\nn\\
&+&\frac 1 2 {\rm tr}[F_{\mu\nu}(V_L)F^{\mu\nu}(V_L)]+
\frac 1 2 {\rm tr}[F_{\mu\nu}(V_R)F^{\mu\nu}(V_R)]
\eea
Notice that, when neglecting the gauge interactions,  the lagrangian is 
invariant under an  extended symmetry corresponding to $(SU(2)_L\otimes
SU(2)_R)^3$. In fact, in this case, we are free to change any of the fields 
$\tilde U$,
$\tilde L$, $\tilde R$ by an independent transformation of a group 
$SU(2)_L\otimes SU(2)_R$ \cite{degene}. As far as the fermions are concerned 
they
transform as in the SM with respect to the group $SU(2)_L\otimes U(1)$.

\resection{The scalar potential}

Let us parameterize the fields
as
\be
\tilde L=\rho_L  L~~~~~ \tilde R= \rho_R  R ~~~~~
\tilde U=\rho_U  U
\ee
with    ${L}^\dagger {L}=I$,   ${R}^\dagger{R}=I$
 and ${U}^\dagger  {U}=I$. 

The scalar potential after these transformations can be rewritten
as
\bea
V(\rho_U,\rho_L,\rho_R)&=& 2 \mu^2 (\rho_L^2+\rho_R^2)+
\lambda (\rho_L^4+\rho_R^4) +2 m^2 \rho_U^2 +
 h \rho_U^4\nn\\ 
&+& 2 f_3 \rho_L^2\rho_R^2 + 2 f \rho_U^2(\rho_L^2+\rho_R^2)
\eea
To study the minimum conditions, let us consider the 
first derivatives of the potential
\be
\frac {\dd \partial V}{\dd\partial\rho_L}=
4 \rho_L (\mu^2 +\lambda \rho_L^2 + f_3 \rho_R^2 + f \rho_U^2)
\ee
\be
\frac {\dd \partial V}{\dd\partial\rho_R}=
4 \rho_R (\mu^2 + \lambda \rho_R^2+ f_3 \rho_L^2+ f \rho_U^2)\ee
\be
\frac {\dd \partial V}{\dd\partial\rho_U}=
4 \rho_U (m^2 + h \rho_U^2 + f (\rho_L^2 + \rho_R^2))
\ee

By considering the vacuum expectation values $<\rho_U>=v$ and
$<\rho_L>=<\rho_R>=u$, the minimum conditions  are
\be
\mu^2+ (f_3+\lambda)u^2 + f v^2=0   
\ee
\be
m^2+ 2fu^2 +h v^2=0   
\ee
from which we get the following solutions
\be
v^2= -\frac {m^2}{h} 
{(1 +\frac {\dd f_3}{\dd \lambda }    
- \frac {\dd 2f\mu^2}{\dd\lambda m^2})}
({\dd 1 +\frac {\dd f_3}{\dd \lambda }    
- \frac {\dd 2f^2}{\dd  h \lambda}})^{-1}
\ee
\be
u^2= -\frac {\mu^2}{\lambda} 
({\dd 1 
- \frac {\dd fm^2}{\dd h \mu^2}})
({\dd 1 +\frac {\dd f_3}{\dd \lambda }    
- \frac {\dd 2f^2}{\dd  h \lambda}})^{-1}
\ee

By considering the second derivatives of the potential
we can get the mass matrix for the three Higgs particles
\be
8 \pmatrix{\lambda u^2&f_3 \uu&f u v\cr
f_3\uu&\lambda \uu&f uv\cr
fuv&fuv&h\vv}
\ee
The mass eigenvalues are
\bea
m_1^2&=&4[(f_3+\lambda)\uu +h \vv -\sqrt{8\uu\vv f^2+
((f_3+\lambda)\uu- h\vv)^2}]\nn\\
m_2^2&=&8\lambda \uu (1- \frac {f_3}{\lambda})\nn\\
m_3^2&=&4[(f_3+\lambda)\uu +h \vv +\sqrt{8\uu\vv f^2+
((f_3+\lambda)\uu- h\vv)^2}]
\eea
Let us comment on the limitations on the
parameters coming from the study of the positivity of
the eigenvalues. 
Adding the requirement of $u^2>0$, $v^2>0$, with the hypothesis $m^2,\mu^2<0$
together with $\lambda,h>0$ for the boundedness of
the potential, we finally get
\be
\lambda -f_3>0,~~~~h>f \frac {m^2}{\mu^2}
\ee
and 
\be
\lambda +f_3> 2 f \frac {\mu^2}{m^2}~~for~f>0~or
\ee
\be
\lambda +f_3> 2 \frac {f^2} {h} ~~for~f<0
\ee

The limit we will be interested in the following is $u\to \infty$ with $v$ 
fixed.
To define the limit one has to look carefully at the minimum conditions (3.6), (3.7). 
It follows from these equations that at least $m^2$ and $\mu^2$ must behave 
like 
$u^2$. Then, in order to keep $v^2$ finite, if follows from eq. (3.8)
that one has also to send $h\to\infty$, unless there is a cancellation
between the leading behaviours of $m^2$ and $\mu^2$ with $u^2$. We can
translate this reasoning in formal terms by requiring that for $u\to \infty$
\be
\mu^2\to au^2,~~~~~~m^2\to bu^2,~~~~~~h\to c\frac{u^2}{v^2}
\ee
The minimum conditions (3.6), (3.7) give the following relations at the
leading order
\be
a=-(f_3+\lambda)~~~~~~b+c=-2f
\ee
We can eventually require that $h$ goes like a constant for
$u\to\infty$, by putting $c=0$ at the end of our calculations. Notice that in 
this
case one has to require $f>0$.
We can now evaluate the leading behaviour of the potential $V$ in terms
of the displaced fields
$\rho_L\to\rho_L+u$, $\rho_R\to\rho_R+u$, $\rho_U\to\rho_U+v$. By neglecting 
a
term independent of the fields we get for $u\to\infty$
\be
V \to 4\lambda u^2(\rho_L^2+\rho_R^2)+8 f_3u^2\rho_L\rho_R+
4c u^2(\rho_U^2+\frac 1 v \rho_U^3+\frac 1{4v^2} \rho_U^4)
\ee
In this limit we can neglect the kinetic term, and the classical
equations of motion are given by
\be
\frac{\partial V}{\partial\rho_L}=\frac{\partial V}{\partial\rho_R}=
\frac{\partial V}{\partial\rho_U}=0
\ee
that is
\be
\rho_L=\rho_R=0
\ee
and
\be
\rho_U(\rho_U^2+3v\rho_U+2v^2)=0
\ee
The last equation has solutions $\rho_U=0,-v, -2v$. The values $\rho_U=0,
-2v$ correspond to two degenerate minima, whereas $\rho_U=-v$ is a
maximum of the potential. However the solution $\rho_U=-2v$ is not a physical 
one 
because it corresponds to a negative value of the original unshifted field, 
which
by hypothesis is positive definite. 
Therefore   the
classical solution is  $\rho_U=0$. We see that in the
case $c\not=0$, the limiting procedure is equivalent to set the fields
$\rho_L$, $\rho_R$ and $\rho_U$ at their minimum. The potential
$V(\rho_L,\rho_R,\rho_U)$ is just a constant, whereas the kinetic term
in (2.8) becomes
\be
{\cal L}^h = \frac 1 4 \lq v^2 Tr (D_\mu U)^\dagger  (D^\mu U)
+u^2 Tr (D_\mu L)^\dagger  (D^\mu L)
+u^2 Tr (D_\mu R)^\dagger  (D^\mu R) \rq            
\ee
In this limit, and at tree level (we are considering the classical
solutions), the linear model described by the lagrangian (2.8)
coincides with the non-linear one discussed in ref \cite{degene},
after identification of the gauge coupling constants
\be
g_0=\tilde g,~~~g_1={\tilde g}',~~~g_2=\frac {g''}{\sqrt{2}}
\ee
and of the parameter
\be
a_2=\frac 1 2 \frac{u^2}{v^2}
\ee
The limit $u\to\infty$ is equivalent to  $a_2\to\infty$.
In this limit we have shown that the heavy vector fields $V_L$ and
$V_R$ decouple and that the non-linear model reduces to the
Higgsless standard model, after the following redefinition of the
gauge couplings
\bea
\frac 1{g^2}&=&\frac 1{{\tilde g}^2}+\frac 2 {{g''}^2}\nn\\
\frac 1{g'^2}&=&\frac 1{{\tilde g}'^2}+\frac 2 {{g''}^2}
\eea
or, in the present notations
\bea
\frac 1 {g^2}&=&\frac 1 {g_0^2}+\frac 1 {g_2^2}\nn\\
\frac 1 {g'^2}&=&\frac 1 {g_1^2}+\frac 1 {g_2^2}
\eea
In this form of the limit it is difficult to argue about the
radiative corrections because the self-coupling of $\rho_U$ is
increasing with $u$. The situation is different when $h$ is finite, that
is $c=0$. In fact, we see from eq. (3.17) that whereas $\rho_L$ and
$\rho_R$ are set to their minimum, the field $\rho_U$ drops out from the
leading term in the potential, and therefore  it is not determined.
So, after setting $\rho_L$ and $\rho_R$ to their minimum we get
an extension of the non-linear model of ref. \cite{degene}, in which
the extra-field $\rho_U$ appears. The potential $V(\rho_L,\rho_R,\rho_U)$ 
coincides with the Higgs field potential 
in the standard model, and ${\cal L}^h$ becomes the standard model Higgs 
kinetic term supplemented by the non-linear pieces in the fields $L$ and $R$.
Therefore, in this case, the limit $u\to\infty$ gives the standard model
with a  Higgs field light with respect to the scale $u$. 

In this
paper we are interested in showing that the linear model, presented
here, satisfies our decoupling requirement 
also at the level of radiative corrections, and
consequently we shall consider  the limit $u\to\infty$
with $c=0$ (that is the self-coupling $h$ fixed).

At the lowest
order in this expansion we get for the Higgs masses

\bea
m_1^2&\sim& 8h \vv \nn\\
m_2^2&\sim&8 \uu (\lambda-  {f_3})\nn\\
m_3^2&\sim& 8u^2(\lambda  +  {f_3})
\eea

\resection{Gauge vector boson spectrum and interactions}    

The  vector boson mass spectrum can be studied in the unitary
gauge $ U=  L=  R =I$ by shifting
the scalar fields as
$\rho_U\to \rho_U +v$, $\rho_{L,R}\to \rho_{L,R} +u$.
We get
\bea
\LL^h&=&\frac 1 2 \left   [ (\dmu \rho_L)^2 + (\dmu \rho_R)^2  +
(\dmu \rho_U)^2 \right ]\nn\\
&+&\frac 1 8  \{ (\rho_L+u)^2  [ g_0^2 (W_3^2 + 2 W^+W^-)
- 2 g_0 g_2 (W_3V_{3L}+ W^-V_{L}^{ +}
+W^+V_{L}^{ -})\nn\\
&+&g_2^2 (V_{3L}^{ 2} + 2 V_L^{ +}V_L^{ -})  ]\nn\\    
&+&(\rho_R+u)^2  [g_1^2 Y^2 -2 g_1 g_2 V_{3R}Y+
g_2^2 (V_{3R}^{ 2} +2 V_R^{+} V_R^{ -})] \nn\\
&+& (\rho_U+v)^2 [ g_0^2 (W_3^2 + 2 W^+W^-) 
-2 g_0 g_1 W_3 Y + g_1^2 Y^2
] \}
\eea
In ref. \cite{degene} we have studied the mass matrices of the vector
bosons in the limit of small ${\tilde g}/g''$, for fixed mass eigenvalues. 
Here we are rather interested in the mass matrices for large mass
eigenvalues of ${V}_{L,R}$, at fixed ${\tilde g}/g''$. 
We will give here only some results of the diagonalization (see \cite{lavoro} 
for
more details).
First of all, it turns out to be convenient to
re-express the results in terms of the parameters $g$ and $g'$ defined in
equation (3.25). In fact, as we have said, these are the relevant
parameters in the limit $u\to\infty$. It is also convenient to introduce
the angle $\varphi$, such that
\be
\sf=\frac g {g_2}=\sqrt{2}\frac g {g''}
\ee
in terms of which
\be
g_0=\frac g {\cf}
\ee
and 
\be
\frac {g_1} {g_0}=\frac {{\tilde g}'}{\tilde g}\equiv\tan{\tilde\theta}=
\frac {\cf \st}{\sqrt{P}}
\ee
where
\be
\tan\theta=\frac {g'} g
\ee
and
\be
P=\ct^2-\sf^2 \st^2
\ee
The mass eigenvalues of the vector bosons are the following:
\hfill\break\medskip\noindent
\underbar{Charged sector}
\hfill\break\medskip\noindent
The fields ${V}^\pm_R$ are unmixed and their mass is given by
\be
M^2_{V^{ \pm}_R}=\frac 1 4 g_2^2 u^2\equiv M^2
\ee
The absence of mixing terms is a consequence of the invariance of the 
lagrangian under
the phase transformation $V^{\pm}_R\to\exp(\pm i\alpha)V^{
\pm}_R$. It will be also convenient to introduce the parameter
\be 
r=\frac 1 4\frac {g^2 v^2}{M^2}=\frac{v^2}{u^2}\frac{g^2}{g_2^2}
\ee
which goes to zero for $u\to\infty$.
The remaining two eigenvalues, in the limit of
small  $r$ are (we keep the same notation $W^\pm$, 
$V^{\pm}_L$ also 
for the mass eigenvectors)
\bea
M^2_{{W^\pm}}&=&\frac \vv 4 g^2 (1-r \sf^2
+\cdots)\nn\\
M^2_{V^{ \pm}_L}&=& \frac \vv 4 g^2 (\frac 1 r \frac 1
{\cf^2}
+\frac {\sf^2}{\cf^2} +r \sf^2
+\cdots)
\eea
Notice that for $r\to 0$, $M^2_{{W^\pm}}$ coincides with the standard
model expression for the $W$ mass. 
\hfill\break\medskip\noindent
\underbar{Neutral sector}
\hfill\break\medskip\noindent
In this sector there is a null eigenvector corresponding to the photon:
\be
\gamma = (s_{\tilde\theta} W_3 +c_{\tilde\theta} Y) \cpsi + \frac {1} 
{\sqrt{2} }
(V_{3L} + V_{3R})\spsi
\ee
where
\be
\tan\psi= \sqrt{2} s_{\tilde\theta} \frac {g_0} {g_2}=\sqrt{2}\frac{\sf\st}
{\sqrt{1- 2\sf^2\st^2}}
\ee
The remaining eigenvalues are, again in the limit of small $r$,
\bea
M^2_{Z}&=&\frac \vv 4  \frac {g^2}{\ct^2}(1-r\sf^2
\frac {1-2\ct^2+2\ct^4}{\ct^4} 
+\cdots)\nn\\
M^2_{V_{3L}}&=& \frac \vv 4 {g^2}
( \frac 1 {r\cf^2} +\frac {\sf^2}{\cf^2}
-r\sf^2 \frac {\ct^2}{1-2\ct^2}+\cdots)\nn\\
M^2_{V_{3R}}&=&\frac \vv 4\frac {g^2}{\ct^2}
(\frac 1{r} \frac {\dd \ct^4}{ P} +\frac {\sf^2\st^4}{P}
+r \frac {\dd\sf^2\st^8}{\dd \ct^4 (1-2\ct^2)}+\cdots)
\eea
Only for small $\varphi$ the heavy vectors are degenerate in mass.

Let us now verify that there are no couplings which increase with $u$ in
the light and in the heavy-light sectors of ${\cal L}^h$. From eq.
(4.1), using the new couplings defined in eq. (3.25) and the diagonalized
vector fields, we get for the light sector of the Higgs-vector interactions
at the leading order in $r$ an expression which coincides with the analogous 
one in the standard model
\be
{\cal L}^h_{\rm light}=\frac {g^2} 4
(\rho_U^2+2\rho_Uv)(W^+W^-+
\frac 1 {2\ct^2}Z^2)
\ee
and for the heavy-light sector
\bea
{\cal L}^h_{\rm heavy-light}&=&\frac {g^2} 4
(\rho_U^2+2\rho_Uv)[-\tan\varphi(W^+V^-_L+W^-V^+_L+\frac 1\ct
ZV_{3L})\nn\\
&+&
\tan^2\varphi~ V^+_LV^-_L+\frac{\sf\tan^2\theta}{\sqrt{P}}ZV_{3R}]
\eea

We will now consider the couplings of the vector bosons to the fermions. 
We assume 
that the fermions have standard transformation properties under the group
$SU(2)_L\otimes U(1)_Y$, and therefore the couplings to the heavy mesons arise
only through the mixing:
\hfill\break\medskip\noindent
\underbar{Charged sector}:
At the first order in $r$ the couplings are given by
\be
{\cal L}_{\rm fermions}^{\rm charged}
=-(a_W W_\mu^-+a_L V^{-}_{L\mu}) J_L^{\mu -}+h.c.
\ee
with
\be
a_W=\frac g {\sqrt {2}}(1-\sf^2 r)
\ee
\be
a_L=-\frac g {\sqrt {2}}(1+\cf^2 r)\tan\varphi
\ee
and $J^\pm_L= \bar \psi_L \gamma^\mu \tau^\pm \psi_L$.
Notice that there is no coupling of $V_{R}^\pm$ to fermions, because
these particles do not mix with the $W^\pm$'s. Also, for $r=0$ the
couplings of $W^\pm$ to the fermions coincide with the standard ones.
\hfill\break\medskip\noindent
\underbar{Neutral sector}:
The couplings are defined by the expression
\bea
{\LL}_{\rm fermions}^{\rm neutral}&=&-e J_{em} \gamma 
 -[AJ_{3L}+B J_{em}]Z\nn\\
&&-[C J_{3L} +D  J_{em}]V_{3L}
-[E J_{3L} +F  J_{em}]V_{3R}
\eea
with 
\be
e=g\st
\ee
 and
\bea
A&=&\frac g\ct (1-\sf^2 \frac {\st^4+\ct^4}{\ct^4}r)\nn\\
B&=& \frac g\ct(-\st^2 +\frac {\sf^2\st^4}{\ct^4}r)\nn\\
C&=&\frac g\ct(-\tan\varphi\ct +\frac {\cf\sf \ct^3}{2\ct^2-1}r)\nn\\
D&=&\frac g\ct\frac {\cf\sf \st^2\ct}{2\ct^2-1}r\nn\\
E&=&\frac g\ct(\frac{\sf\st^2}{\dd\sqrt{P}}+\frac {\dd\sf\st^6\sqrt{P}}{\dd\ct
(1-2\ct^2)}r)\nn\\
F&=&\frac g\ct(-\frac{\sf\st^2}{\dd\sqrt{P}}-\frac {\dd\sf\st^4\sqrt{P}}
{\dd\ct^4}r)
\eea

The expression for the electric charge is valid  to all order in $r$, 
while the other coefficients in (4.20) are given only at first order in $r$.
In particular the couplings to the $Z$ go back to their standard model
values for $r\to 0$. Again, there are no couplings increasing when
$r\to 0$, both in the charged and in the neutral sector.

Finally we have to examine the gauge boson self-couplings. 
Let us define the following formal combination
\be
AB^-C^+=A^{\mu\nu}B_\mu^-C_\nu^++A^\nu(B_{\mu\nu}^-C^{\mu +}-
B_{\mu\nu}^+C^{\mu -})
\ee
where
\be
A^{\mu\nu}=\partial^\mu A^\nu-\partial^\nu A^\mu
\ee
and similar expression for $B_{\mu\nu}^\pm$. Then 
the trilinear gauge boson couplings in terms of the original fields are
given by
\be
\LL =i[g_0 W_3 W^- W^+ 
+g_2
V_{3L} V^{-}_{L} V^+_{L}
+g_2
V_{3R} V^{-}_{R} V^+_{R}] 
\ee
Using the redefinition of the couplings and the expressions for the mass
eigenstates, we find, again at the first order in $r$, for the
light sector
\be
{\cal L}_{\rm light}=ig[\st \gamma W^-W^++\ct Z W^-W^+]
\ee
and for the heavy-light sector
\bea
{\cal L}_{\rm heavy-light}&=&ig [\st \gamma ({V}_L^-{V}_ L^+ +{V}_R^-{V}_ R^+)
+(\ct +r\frac 1 \ct (2
\cf^2-1))Z {V}_L^-{V}_L^+\nn\\
 &+&\cf \sf \frac
{r}{\ct}(ZW^-{V}_L^++Z{V}_L^- W^+)-\frac {\st^2}{ \ct} (1+r \frac {1} {\ct^4}
P)Z{V}_R^-{V}_R^+\nn\\
&+&(1-r(1-2\cf^2))({V}_{3L} W^-{V}_L^+ +{V}_{3L}{V}_L^-W^+)
+r\cf\sf  {V}_{3L}W^- W^+\nn\\
&-&r\frac{\cf\st^2\sqrt{P}}{\ct(1-2\ct^2)}({V}_{3R}W^-{V}_L^+ +
{V}_{3R}{V}_L^-W^+)\nn\\
&+&r\frac{\sf\st^2\sqrt{P}}{\ct^3}{V}_{3R}W^- W^+]
\eea

The quadrilinear couplings are obtained starting from
\bea
&-{\dd \frac { g_0^2} 2} S_{\mu\nu\rho\sigma}& 
[W_\mu^+ W_\nu^- ( W_\rho^+ W_\sigma^- + W_{3\rho} W_{3\sigma})\nn\\
&&+\frac 1 {\tan^2\varphi}
V_{L\mu}^+ V_{L\nu}^- ( V_{L\rho}^+ V_{L\sigma}^- + V_{3L\rho}
V_{3L\sigma})\nn\\
&&+\frac 1 {\tan^2\varphi}
V_{R\mu}^+ V_{R\nu}^- ( V_{R\rho}^+ V_{R\sigma}^- + V_{3R\rho}
V_{3R\sigma})]
\eea
with $S_{\mu\nu\rho\sigma}=2g_{\mu\nu} g_{\rho\sigma}
-g_{\mu\rho} g_{\nu\sigma} -g_{\mu\sigma} g_{\nu\rho}$.

At the lowest order in $r$
one gets for the light part (in this case the corrections are of the order 
$r^2$)
\bea
{\cal L}_{\rm light}=&- {\dd\frac {  g^2}  2} S^{\mu\nu\rho\sigma} &
[W_\mu^+ W_\nu^- ( W_\rho^+ W_\sigma^-+\ct^2 Z_\rho Z_\sigma\nn\\ 
&&+2\ct\st \gamma_\rho Z_\sigma 
+\st^2 \gamma_\rho\gamma_\sigma)]
\eea
and for the heavy-light part (here the corrections are of order $r$)
\bea
{\cal L}_{\rm heavy-light}=&-{\dd \frac {  g^2}  2} S^{\mu\nu\rho\sigma} &
[W_\mu^+ W_\nu^- (
{V}_{3L\rho}{V}_{3L\sigma}+{V}_{L\rho}^+ {V}_{L\sigma}^-)\nn\\
&&+(W_\mu^+{V}_{L\nu}^-+{V}_{L\mu}^+ W_\nu^-) ({V}_{L\rho}^+ W_\sigma^- + 
W_\rho^+ {V}_{L\sigma}^-\nn\\
&&+\frac{2\cf^2-1}{\cf\sf}({V}_{L\rho}^+{V}_{L\sigma}^- +
{V}_{3L\rho}{V}_{3L\sigma}))\nn\\
&&+{V}_{L\mu}^+ {V}_{L\nu}^- ( W_\rho^+ W_\sigma^- +\ct^2  Z_\rho Z_\sigma 
+2\ct\st \gamma_\rho Z_\sigma  + \st^2 \gamma_\rho\gamma_\sigma\nn\\
&&+\frac{2\cf^2-1}{\cf\sf}({V}_{L\rho}^+{V}_{L\sigma}^-+
{V}_{L\rho}^+{V}_{L\sigma}^-\nn\\
&&+2\ct {V}_{3L\rho}{Z}_{\sigma}+2\st {V}_{3L\rho}{\gamma}_{\sigma}))\nn\\
&&+{V}_{R\mu}^+{V}_{R\nu}^-(\frac{\st^4}{\ct^4}Z_\rho Z_\sigma-
2\frac{\st^3}{\ct}\gamma_\rho Z_\sigma+\st^2\gamma_\rho\gamma_\sigma\nn\\
&&-2\frac{\st\sqrt{P}}{\sf\ct^2}(\st {V}_{3R\rho}{Z}_{\sigma}-
\ct{V}_{3R\rho}{\gamma}_{\sigma})]
\eea
Both the trilinear and quadrilinear light parts of the lagrangian agree with 
the 
standard model results, and the heavy-light sectors do not show any coupling 
increasing 
with the heavy mass $M$.

\resection{Conclusions}

In this work we have formulated a renormalizable model which can be suitable 
to 
describe at some intermediate energy a scenario of the kind considered
in the non-commuting extended technicolor schemes. The gauge symmetries
of the model are an extension of the SM symmetries by an extra
$SU(2)'_L\otimes SU(2)'_R$ factor. In the limit in which the expectation
value of the Higgs fields related to the new symmetries gets very large,
one recovers a previously considered non-linear model \cite{degene}. 
The main property
of the non-linear model was its decoupling for large values of the masses
of the gauge bosons associated to the extra-symmetry factors, in spite
of its non-linearity. The masses of the fields $V_L$ and $V_R$ go to
infinity together with the expectation values of the related 
the Higgs fields, and therefore one recovers in this limit the
non-linear realization of the standard model, or the Higgsless standard
model. 

In the present paper we have also considered a slight modification of
this limiting procedure which allows to the normal Higgs field (the one
associated to the global symmetry $SU(2)_L\otimes SU(2)_R$) to remain
light (with respect to the heavy scale). We can then decompose the 
lagrangian into three pieces: the light sector, involving the SM fields 
including the light Higgs, the heavy-light and the heavy sectors. The
light sector part of the lagrangian is identical to the lagrangian of
the SM, and the heavy-light part does not contain coupling increasing
with the heavy scale. Having isolated all the big parameters in the
heavy part, one can conclude that at tree level there is decoupling
(this has been shown explicitly for the case of the non-linear model in ref.
\cite{degene}).
Furthermore, by the arguments leading to the Appelquist-Carazzone
\cite{AC} theorem, one can argue that the decoupling must hold also at
the level of the radiative corrections. An explicit proof of this
statement will be given in a more technical paper \cite {lavoro}. As a
consequence, in the low-energy region the model cannot be distinguished
by the SM. However, as shown in the case of the non-linear model 
\cite {degene}, 
when approaching the
threshold for the production of the heavy vector states it is possible to have
big deviations, and more spectacular effects after the threshold.

\begin{center}
\begin{bf} 
 ACKNOWLEDGMENTS
  \end{bf}
\end{center}

We would link to thank Prof. R. Gatto for enlightening discussions.

This work is part of the EEC Project "Tests of electroweak symmetry breaking 
and future European colliders", No. CHRXCT94/0579.


\begin{thebibliography}{99}

\bibitem{degene}
R.Casalbuoni,  D.Dominici, A. Deandrea, R.Gatto, S.De Curtis and M. Grazzini,
    Phys. Rev. {\bf D53}  (1996) 5201  
\bibitem{nctc}
R.S.Chivukula, E.H.Simmons and J. Terning,
     Phys. Rev. {\bf D53}  (1996) 5258 
\bibitem{lavoro} 
R.Casalbuoni,  D.Dominici, S.De Curtis and M. Grazzini,
to be published
\bibitem{AC} 
T.Appelquist and J.Carazzone, Phys. Rev.  {\bf D11} (1975) 2856
\end{thebibliography}
\end{document}